\documentclass[aps,prl,twocolumn,showpacs,groupedaddress,floatfix]{revtex4}

\usepackage{graphicx}
\usepackage{amsmath,amssymb}
\usepackage{times,txfonts}

\newcommand{\ket}[1]{\ensuremath{\,|{#1}\rangle}}

\newcommand{\matrixe}[3]{\ensuremath{\langle{#1}|\,{#2}\,|{#3}\rangle}}

\begin{document}

\title{Correlation energies in the random phase approximation using realistic interactions}

\author{C. Barbieri}
\affiliation{Gesellschaft f\"ur Schwerionenforschung, Planckstr. 1,  D-64291, Darmstadt, Germany}

\author{N. Paar, R. Roth, P. Papakonstantinou}
\affiliation{Institut f\" ur Kernphysik, Technische Universit\" at Darmstadt, Schlossgartenstr. 9,
D-64289 Darmstadt, Germany}

\date{\today}

\begin{abstract}
The self-consistent random phase approximation (RPA) based on a correlated realistic nucleon-nucleon interaction is used to evaluate correlation energies in closed-shell nuclei beyond the Hartree-Fock level. The relevance of contributions associated with charge exchange excitations as well as the necessity to correct for the double counting of the second order contribution to the RPA ring summation are emphasized. Once these effects are properly accounted for, the RPA ring summation provides an efficient tool to assess the impact of long-range correlations on binding energies throughout the whole nuclear chart, which is of particular importance when starting from realistic interactions.
\end{abstract}

\maketitle


On the mean-field level, the nuclear many-body problem can be treated in the well known Hartree-Fock~(HF) approximation. In conjunction with phenomenological interactions, this scheme is successful in describing various nuclear ground state properties. Recently, realistic nucleon-nucleon~(NN) interactions have also been regulated for implementation in nuclear structure calculations in two novel approaches, (i) low-momentum NN interaction from the renormalization group theory, $V_{\text{low}k}$~\cite{Bog.03}, and (ii) correlated interactions $V_{\text{UCOM}}$ constructed in the framework of unitary correlation operator method~\cite{Rot.04}. 

In connection with these interactions, the HF approximation underestimates the binding energy due to the inadequate representation of long-range correlations. In recent HF studies based on $V_{\text{low}k}$ and $V_{\text{UCOM}}$, these missing correlations have been recovered within many-body perturbation theory (MBPT) by evaluating corrections up to third order \cite{Cor.05,Rot.06}. In general, correlations beyond HF are more relevant in studies based on realistic NN interactions than in phenomenological models based on, e.g., Skyrme or Gogny functionals, because the latter already mimic part of the many-body correlations through the phenomenological fit. Conversely, studies based on realistic interactions allow for more profound insights into the many-body dynamics~\cite{Dic.04}.

Going beyond low-order MBPT, correlation energies can be obtained using the random phase approximation (RPA) as a tool to evaluate a partial summation over particle-hole ring diagrams. In the language of RPA, ground state correlations emerge from a coupling to giant resonances and surface vibrations~\cite{Fuk.64,Row.68,Don.00,Shi.00}. The correlation effects of these two classes of collective motion are different---giant resonances influence the binding energies, while surface modes have a more pronounced effect on charge density distributions~\cite{Rei.85,Esb.83}. By applying the RPA method, these excitations are approximated as harmonic vibrations. The total binding energy can then be evaluated as the sum of the zero-point energies of all the possible modes. The RPA framework has been employed in several studies of the ground state correlation energies mainly associated with quadrupole and octupole modes, and pairing vibrations~\cite{Ula.69,Fri.86,McN.90,Ste.02,Bar.04}. It should be noted that for systems made of more than one fermion species, other modes, e.g. Gamow-Teller transitions in nuclei, become possible. However, no discussion is usually found in the literature regarding the relevance of charge exchange excitations. 

In this work we study RPA correlation energies in conjunction with correlated realistic NN interactions derived from the Argonne V18 potential~\cite{Wir.95} in the framework of the unitary correlation operator method (UCOM). The short-range central and tensor correlations induced by the realistic potential are described by a unitary state-independent transformation \cite{Rot.04,Rot.05}. The central correlations induced by the strong short-range repulsion (core) of the interaction are described by a unitary shift in the relative coordinate of all nucleon pairs~\cite{Rot.04,Rot.05}. The tensor part of the interaction induces strong correlations between spin and relative orientation of the nucleons. Only the short-range system-independent part of these tensor correlations is described explicitly by the unitary transformation, the effect of long-range tensor correlations as well as all residual correlations has to be described by the many-body method employed.

In this framework we obtain a correlated interaction $V_{\text{UCOM}}$ which is phase-shift equivalent to, but much softer than the original potential. Technically, the unitary transformation modifies the off-shell behavior of the interaction by introducing strong non-local terms. The application of $V_{\text{UCOM}}$ in no-core shell model calculations for $^3$H and $^4$He shows a dramatic improvement of the convergence behavior \cite{Rot.05}. At the same time, the tensor correlator can be tuned in order to minimize the effect of the net three-body force, which is the sum of the genuine three-body potential accompanying the bare NN potential and the three-body terms induced by the unitary transformation of the Hamiltonian. As no-core shell model calculations show, the two-body $V_{\text{UCOM}}$ provides a good quantitative description of ground and low-lying excited states throughout the p-shell \cite{NaRo06}. Using the same correlated interaction $V_{\text{UCOM}}$ we have performed HF and RPA calculations for closed shell nuclei throughout the nuclear chart~\cite{Rot.06,Paa.06}. Based on the HF solution the effect of long-range correlations on the binding energies was estimated via MBPT and good agreement with experimental binding energies was found \cite{Rot.06}. 

In the present work we investigate the effect of long-range correlations in the framework of a fully self-consistent RPA~\cite{Paa.06}. Using the single-particle basis resulting from HF, the RPA configuration space is built and the generalized eigenvalue problem posed by the RPA equations is solved,
\begin{equation}
\label{rpaeq}
\left(
\begin{array}{cc}
A & B \\
B^{^\ast} & A^{^\ast}
\end{array}
\right)
\left(
\begin{array}{c}
X^{\nu} \\
Y^{\nu}
\end{array}
\right) =\omega_{\nu}\left( 
\begin{array}{cc}
1 & 0 \\
0 & -1
\end{array}
\right)
\left( 
\begin{array}{c}
X^{\nu} \\
Y^{\nu}
\end{array}
\right)\; ,
\end{equation}
where the eigenvalues $\omega_{\nu}$ correspond to RPA excitation energies. The forward and backward-going particle-hole amplitudes, $X_{ph}^\nu$ and $Y_{ph}^\nu$, respectively, are related to the transition amplitudes to the excited states $\ket{\Psi_\nu}$ by
\begin{equation}
\label{Zab}
\matrixe{\Psi_\nu}{c^\dag_\alpha c_\beta}{\Psi_0} ~=~
\delta_{\alpha,p} \delta_{\beta,h} \;  X_{ph}^\nu ~+~
\delta_{\alpha,h} \delta_{\beta,p} \;  Y_{ph}^\nu \; .
\end{equation}
Here and in the following, the indices $p$~($h$) always label particle (hole) states while Greek letters refer to any orbits. $c^\dag$ ($c$) are the usual creation (annihilation) operators. The HF+RPA scheme is applied in a fully self-consistent way, i.e. the same translational invariant Hamiltonian $H_{\text{int}} = T - T_{\text{cm}} + V_{\text{UCOM}}$, which formally is a two-body operator, is used in the HF equations that determine the single-particle basis and in the RPA matrices $A$ and $B$. This ensures that the RPA amplitudes do not contain spurious components associated with the center-of-mass translational motion~\cite{Paa.06}.

There are several approaches to calculate the RPA ground-state correlation energy. In the present study, we consider two different formulations, based on (A) the direct evaluation of the expectation value of the Hamiltonian and (B) the quasi-boson approximation. The expectation value of the Hamiltonian $H_{\text{int}}$ (or, in general, the two-body part of the Hamiltonian) is given by
\begin{eqnarray}
\lefteqn{ \matrixe{\Psi_0}{H_{\text{int}}}{\Psi_0} =
       \frac{1}{4} \sum_{\alpha \beta \gamma \delta} H_{\alpha \beta, \gamma \delta}
       \matrixe{\Psi_0}{c_\alpha^\dag c_\beta^\dag c_\delta c_\gamma}{\Psi_0}
       }
&&
\nonumber\\
 &=& 
 \frac{1}{4} \sum_{\alpha \beta \gamma \delta} H_{\alpha \beta, \gamma \delta}
   \matrixe{\Psi_0}{\delta_{\beta\delta} c_\alpha^\dag c_\gamma -
   c_\alpha^\dag c_\delta  c_\beta^\dag c_\gamma}{\Psi_0}
\nonumber\\
 &=& 
 \frac{1}{4} \sum_{\alpha \beta \gamma \delta} H_{\alpha \beta, \gamma \delta}
   \left[
    \delta_{\beta\delta} \matrixe{\Psi_0}{c_\alpha^\dag c_\gamma}{\Psi_0}
   \right.
\nonumber\\
 &&\hspace{1.5cm}-
   \left.
    \matrixe{\Psi_0}{c_\alpha^\dag c_\delta}{\Psi_0}
    \matrixe{\Psi_0}{c_\beta^\dag  c_\gamma}{\Psi_0} 
   \right]
\label{Vexpval}\\
 & &- 
 \frac{1}{4} \sum_{\alpha \beta \gamma \delta} H_{\alpha \beta, \gamma \delta}
    \sum_\nu
    \matrixe{\Psi_0}{c_\alpha^\dag c_\delta}{\Psi_\nu}
    \matrixe{\Psi_\nu}{c_\beta^\dag  c_\gamma}{\Psi_0} \;,
\nonumber 
\end{eqnarray}
where $H_{\alpha \beta, \gamma \delta}$ are the antisymmetrized matrix elements of $H_{\text{int}}$. Eq.~(\ref{Vexpval}) is still exact. By approximating the transition amplitudes according to Eq.~(\ref{Zab}) and employing the RPA equations~(\ref{rpaeq}) one is led to express the total energy of the system as
\begin{equation}
E = E_{\text{HF}} + E_{\text{RPA}},
\end{equation}
where the HF binding energy supplemented with the RPA correlation energy,
\begin{equation}
E_{\text{RPA}}^{\text{(A)}} = -\frac{1}{2}\sum_{\nu} \sum_{p \; h}
  \left({\epsilon}_p - {\epsilon}_h + \hbar \omega_{\nu} \right)
  |Y^{\nu}_{ph}|^2 .
\label{barbform}
\end{equation}

In the quasi-boson approximation the ground state energy is described as the zero point energy of a collection of harmonic vibrations. Using the oscillator-projection method by Rowe~\cite{Row.68} one obtains,
\begin{equation}
E_{\text{RPA}}^{\text{(B)}} = -\sum_{\nu} \hbar {\omega}_{\nu}\sum_{p \; h}|Y^{\nu}_{ph}|^2 \; .
\label{roweform}
\end{equation}
Eqs.~(\ref{barbform}) and~(\ref{roweform}) employ the RPA eigenvalues $\omega_{\nu}$ and backward-going amplitudes $Y^{\nu}_{ph}$ as well as the HF single-particle levels ${\epsilon}_\alpha$. From the completeness insertion in the last row of Eq.~(\ref{Vexpval}) and the summation over all (proton and neutron) orbits, it is clear that one needs to sum over all possible final states predicted by the RPA formalism, independently of isospin. Hence, the sum over $\nu$ in Eqs.~(\ref{barbform}) and~(\ref{roweform}) includes not only all the multipolarities and parities but {\em also} charge-exchange processes. 

In comparison to a order-by-order summation of particle-hole ring diagrams in MBPT, the above formulae implicitly double count the second order contribution, as pointed out by Fukuda et al.~\cite{Fuk.64} and Ellis \cite{Ell.70.87}. To show this, we expand the RPA eigenvalues $\omega_{\nu}$ and the amplitudes $X^{\nu}_{ph}$ and $Y^{\nu}_{ph}$ in a perturbation series of the interaction $H_{\text{int}}$, as done in  Ref.~\cite{Fuk.64}. Inserting these into Eqs.~(\ref{barbform}) and~(\ref{roweform}) one obtains, 
\begin{eqnarray}
E_{\text{RPA}}^{\text{(A)}} 
  &=& 2 E^{(2)} + \frac{3}{2} E^{(3)}_{\text{ring}} + {\cal O}(H^4)
\label{barbPT}
\\
E_{\text{RPA}}^{\text{(B)}} 
  &=& 2 E^{(2)} + E^{(3)}_{\text{ring}} + {\cal O}(H^4) \; ,
\label{rowePT}
\end{eqnarray}
where $E^{(2)}$ corresponds to the second order contribution in MBPT
\begin{equation}
  E^{(2)} = -\frac{1}{4} \sum_{p_1 p_2 h_1 h_2}
 \frac{H_{p_1 p_2, h_1 h_2} \; H_{h_1 h_2, p_1 p_2}}
     {{\epsilon}_{p_1} + {\epsilon}_{p_2} - {\epsilon}_{h_1} - {\epsilon}_{h_2} }
\; ,
\label{PT2}
\end{equation}
and $E^{(3)}_{\text{ring}}$ to the contribution of the ring diagram at third order
\begin{eqnarray}
  E^{(3)}_{\text{ring}} &=& \sum_{p_1 p_2 p_3} ~ \sum_{h_1 h_2 h_3}
\label{PT3r}  \\
 && \frac{H_{p_1 p_2, h_1 h_2} \;  H_{p_3 h_2, h_3 p_2} \;  H_{h_1 h_3, p_1 p_3}}
     {({\epsilon}_{p_1} + {\epsilon}_{p_2} - {\epsilon}_{h_1} - {\epsilon}_{h_2})
      ({\epsilon}_{p_1} + {\epsilon}_{p_3} - {\epsilon}_{h_1} - {\epsilon}_{h_3}) }
\; .
\nonumber
\end{eqnarray}
The double counting of the second order contribution $E^{(2)}$ is evident and has to be corrected for explicitly. Beyond the second order, $E_{\text{RPA}}^{\text{(B)}}$ does not introduce any further overcountings~\cite{Fuk.64}, while $E_{\text{RPA}}^{\text{(A)}}$ remains somewhat troublesome. The double counting of $E^{(2)}$ is intrinsic to the quasi-boson approximation and can be avoided only in a formalism that (beyond the RPA approach) explicitly recouples particle and hole states between different phonons. This can be achieved at the level of the many-body self-energy~\cite{Bar.01.02}.

One has to keep in mind that the particle-hole ring summation does not include all possible diagrams, e.g., the third order ring term $E^{(3)}_{\text{ring}}$ is only one of three third order contributions and two-particle two-hole diagrams are neglected by both Eq.~(\ref{barbform}) and~(\ref{roweform}). The latter are known to be approximately compensated by Pauli exchange effects between different phonons at high order~\cite{Fes.74}. 

For a microscopic theory based on realistic nucleon-nucleon interactions, the correlation energy beyond the simple HF approximation is sizable. This remains true even after regularizing the interaction to account for the effects of short-range correlations. The residual long-range correlations manifest themselves in a sizable second-order contribution, $E^{(2)}$ \cite{Rot.06}. Therefore, the naive application of Eqs.~(\ref{barbform}) and~(\ref{roweform}) would lead to a strong overestimation of the correlation energy. In the context of phenomenological models Eq.~(\ref{roweform}) is often applied without corrections for double countings \cite{Ula.69,Fri.86,McN.90,Ste.02,Bar.04}. This might be compensated by also neglecting the contribution from charge exchange terms. However, these are two (not well controlled) errors that do not necessarily cancel each other \cite{Paa.proc.05}.

\begin{figure}
\includegraphics[width=\columnwidth,clip=true]{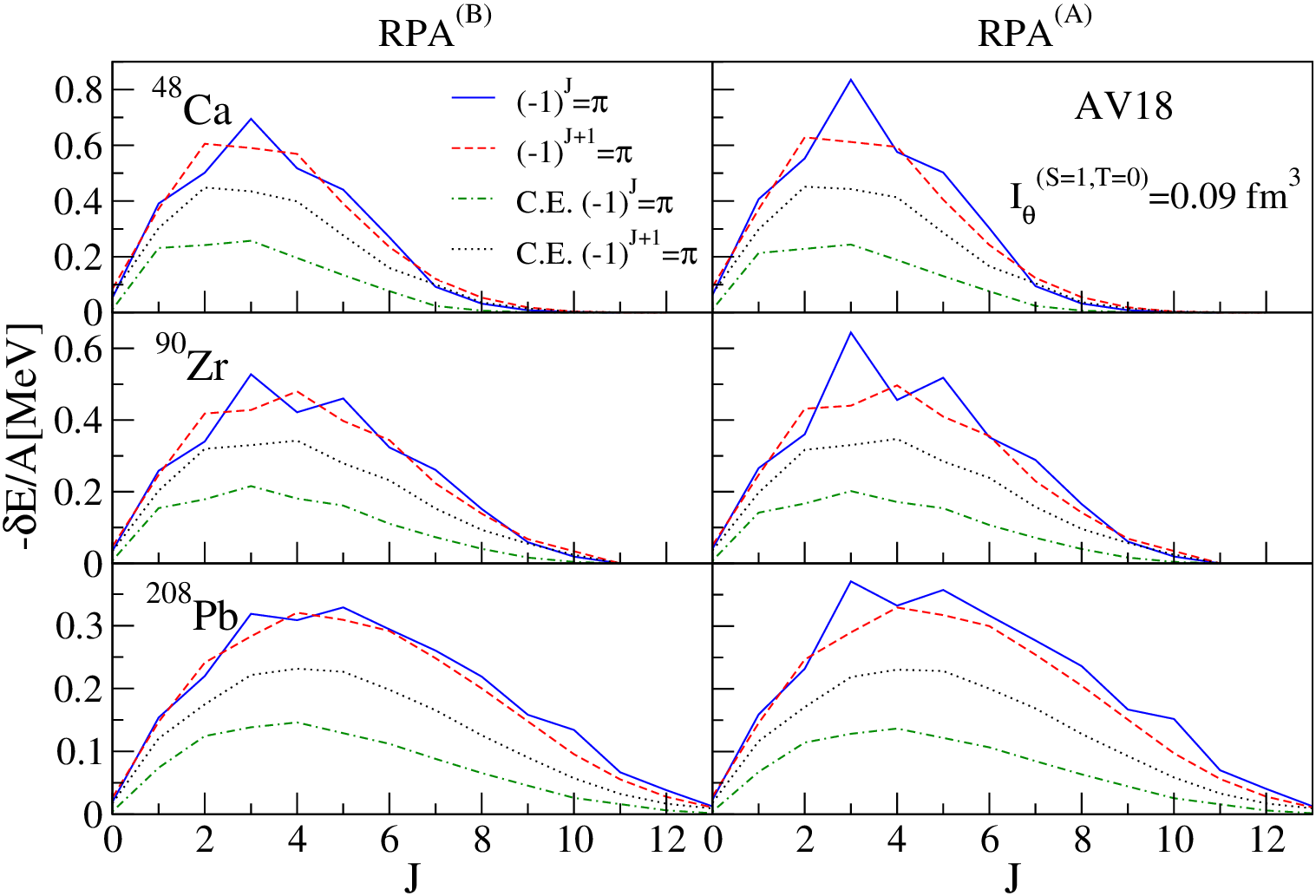}
\caption{(Color online) Partial contributions to the RPA correlation energies in $^{40}$Ca, $^{90}$Zr, and $^{208}$Pb as a function of multipolarity $J$. The correlation energies are evaluated using Eq.~(\ref{roweform}) (left panels) and Eq.~(\ref{barbform}) (right panels). Solid and dashed lines correspond to contributions from natural and unnatural parity excitations, respectively. Contributions from charge-exchange excitations (C.E.) for natural and unnatural parities are represented by dot-dashed and dotted lines.}
\label{figcorren1}
\end{figure}

We apply this scheme to evaluate correlation energies for closed-shell nuclei throughout the nuclear chart based on a realistic NN interaction. We employ the $V_{\text{UCOM}}$ derived from the Argonne V18 interaction using the optimal correlation operators determined in Ref. \cite{Rot.05}. The range of the tensor correlator in the triplet-even channel was fixed to reproduce the binding energies of $A\leq4$ nuclei in no-core shell model calculations ($I_{\vartheta}=0.09\,\text{fm}^3$). The same correlator was used successfully in HF and MBPT calculations reported in Ref. \cite{Rot.06}. For a systematic calculation of the RPA correlations, Eq.~(\ref{rpaeq}) was projected onto good angular momentum $J$ and parity $\pi$ and all available multipolarities were taken into account. All calculations were performed using $13$ major harmonic oscillator shells. The spurious contributions in the $1^-$ channel were excluded from the calculations of correlation energies.

\begin{figure}
\includegraphics[width=\columnwidth,clip=true]{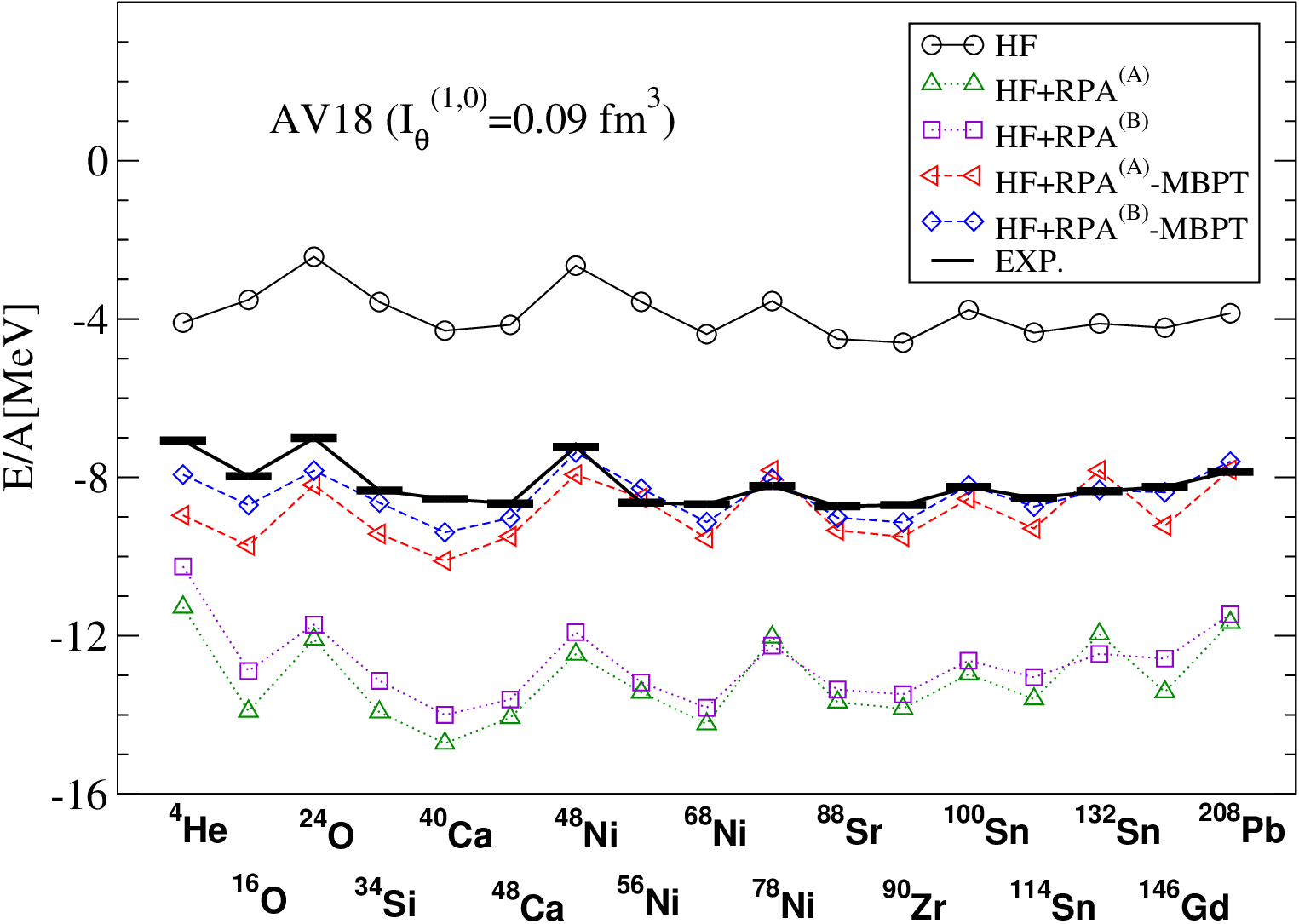}
\caption{(Color online) Binding energies per nucleon for a series of closed-shell nuclei in comparison with experiment. Shown are the HF energy, the energy including RPA correlations according to Eqs.~(\ref{barbform}) and (\ref{roweform}) (labeled HF+RPA${}^{\text{(A)}}$ and HF+RPA${}^{\text{(B)}}$, resp.), and those including RPA correlations corrected for the double counting of the second order (denoted HF+RPA${}^{\text{(A,B)}}$-MBPT).
All calculations are based on the correlated Argonne V18 interaction.}
\label{figcorren2}
\end{figure}

In Fig.~\ref{figcorren1} we display the individual contributions to the correlation energies evaluated in the two different formulations (Eqs.~(\ref{barbform}) and (\ref{roweform})) for $^{48}$Ca, $^{90}$Zr, and $^{208}$Pb. Shown are the contributions to $E_{\text{RPA}}^{\text{(A)}}$ and $E_{\text{RPA}}^{\text{(B)}}$ as function of the multipolarity $J^{\pi}=0^{\pm}-13^{\pm}$ separated into natural parity, $\pi=(-1)^J$, and unnatural parity, $\pi=(-1)^{J+1}$, excitations as well as charge exchange excitations. In general, the RPA correlation energy increases with $J$, reaches the maximum for $J=3-4$, and slowly decreases towards higher multipolarities. Both, natural and unnatural parity states are equally important and charge-exchange excitations also have significant contributions to the correlation energy. Although in all nuclei the largest contributions come from the $J=3-4$ excited states, one obviously needs to include all other multipolarities as well. This is especially important for heavier nuclei, where the correlation energies are widely distributed over various multipolarities up to $J=13$.

The overall sum of correlation energies displayed in Fig.~\ref{figcorren1} provides the correction to the binding energies in finite nuclei. In Fig.~\ref{figcorren2} we show binding energies per nucleon for several closed-shell nuclei obtained in HF with and without the inclusion of the RPA correlation energies, in comparison to the experimental binding energies~\cite{Aud.95}. The plain HF calculations underestimate the binding energies due to the inadequate description of long-range correlations. Inclusion of the correlation energies resulting from Eqs. (\ref{barbform}) and (\ref{roweform}) without correction for the double-counting of the second order contribution leads to a strong overbinding. As discussed by da Providencia~\cite{Pro.68}, only after explicit correction for the double counting by subtracting the second order contribution, i.e. by using $E_{\text{HF}}+E_{\text{RPA}}-E^{(2)}$, we obtain a proper estimate for the binding energy including correlation effects which is in good agreement with experiment. The difference between the two schemes for the evaluation of the RPA correlation energy traces back to the overcountings of higher-order terms in Eq. (\ref{barbform}). Only $E_{\text{RPA}}^{\text{(B)}}$ resulting from Eq. (\ref{roweform}) corresponds directly to the particle-hole ring summation.

\begin{figure}
\includegraphics[width=\columnwidth,clip=true]{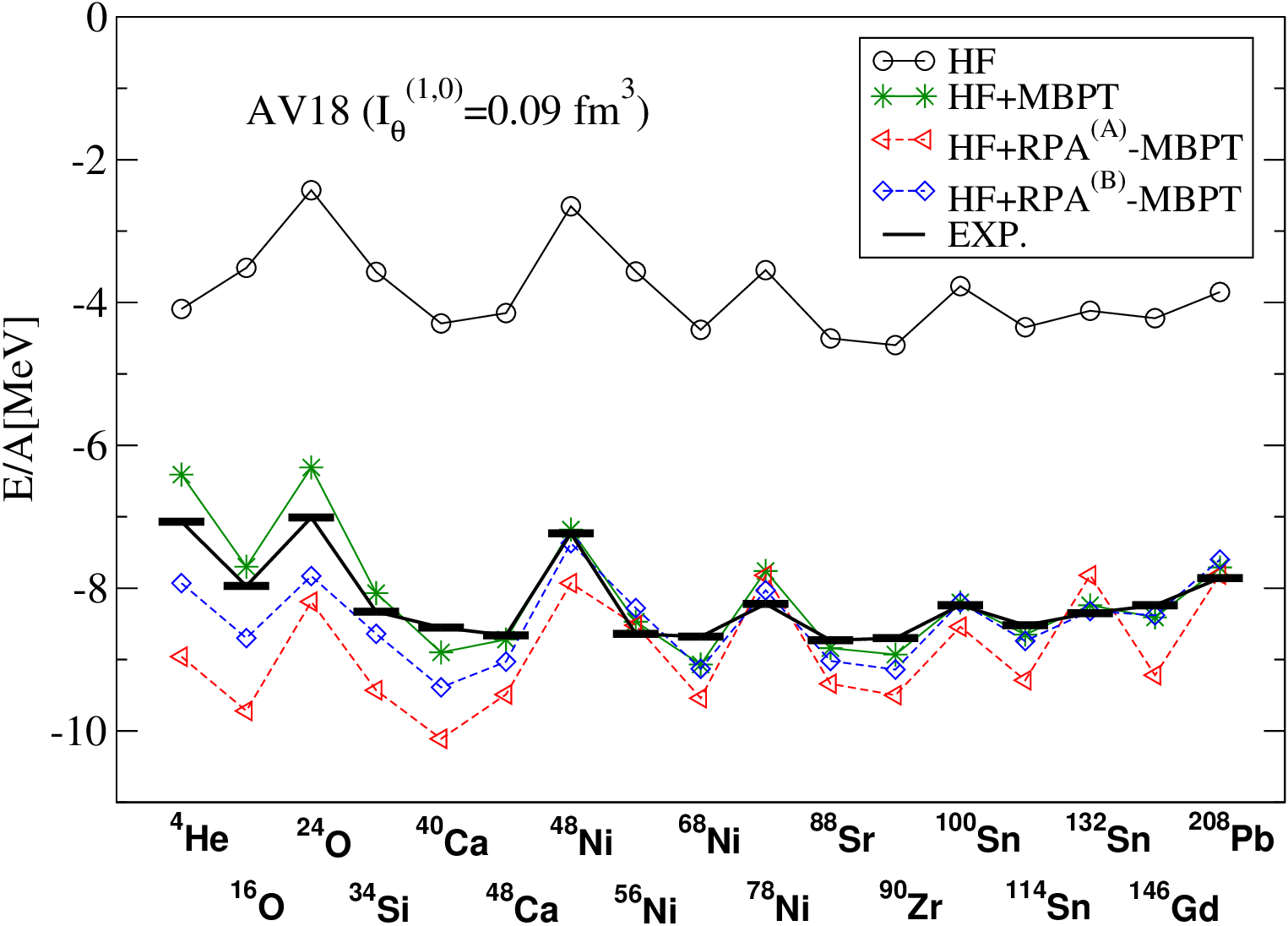}
\caption{(Color online) Binding energies per nucleon including RPA correlations and second-order correction  (HF+RPA${}^{\text{(A,B)}}$-MBPT) compared to energies resulting from second-order perturbation theory only (labeled HF+MBPT). Compare to Fig.~\ref{figcorren2}.} 
\label{figcorren3}
\end{figure}

For a more detailed discussion, in Fig.~\ref{figcorren3} we show the corrected RPA energies, $E_{\text{HF}}+E_{\text{RPA}}-E^{(2)}$, in comparison to the direct second order perturbative estimate, $E_{\text{HF}}+E^{(2)}$. The binding energies per nucleon obtained by employing the second order perturbation theory agree rather well with the results of the RPA ring summation. Although the second-order correction to the HF energy is large, the higher order contributions included in the ring summation seem to have a relatively small net effect. Beyond ${}^{48}$Ca the binding energies per nucleon including $E_{\text{RPA}}^{\text{(B)}}$ are in excellent agreement with the perturbative second-order results and also with experimental values. The ring summation provides systematically larger correlation energies than the plain second-order. This is in line with no-core shell model calculations for light isotopes which also predict energies somewhat lower than the second-order estimate \cite{NaRo06}.  

In conclusion, we have employed a self-consistent RPA approach to evaluate correlation energies based on a correlated realistic NN-potential. Correlations beyond the HF level have a sizable impact even if one uses regularized interactions like $V_{\text{UCOM}}$. We point out the need to sum over all possible excitation modes, including charge-exchange excitations, and to correct for the double counting of the second-order contribution, when using standard expressions like Eq.~(\ref{roweform}) to evaluate the correlation energy. Then the RPA ring summation provides an efficient tool to evaluate correlation energies throughout the nuclear chart. In connection with $V_{\text{UCOM}}$, the RPA correlation energies generally confirm the results of second order MBPT, indicating that the net contribution of higher-order ring-diagrams are moderate although the second order itself is large. Both, RPA ring summation and low-order MBPT, provide efficient tools for nuclear structure calculations with correlated realistic NN-interactions.

\acknowledgments
This work is supported by the Deutsche Forschungsgemeinschaft (DFG) under contract SFB 634.


\end{document}